\documentclass[12pt,a4paper,epsf]{article}
\usepackage{graphics}
\usepackage{amssymb,amsmath}
\usepackage[dvips]{lscape,graphicx}
\usepackage{cite}
\usepackage{longtable}
\usepackage{slashed}
\usepackage{bbold}

\newcommand{\ct}{\cite}
\newcommand{\lb}{\label}

\newcommand{\bc}{\begin{center}}
\newcommand{\ec}{\end{center}}
\newcommand{\bd}{\begin{displaymath}}
\newcommand{\ed}{\end{displaymath}}
\newcommand{\be}{\begin{equation}}
\newcommand{\ee}{\end{equation}}
\newcommand{\ba}{\begin{array}}
\newcommand{\ea}{\end{array}}
\newcommand{\bea}{\begin{eqnarray}}
\newcommand{\eea}{\end{eqnarray}}
\newcommand{\bt}{\begin{tabular}}
\newcommand{\et}{\end{tabular}}
\newcommand{\un}{\underline}

\newcommand{\bp}{\begin{picture}}
\newcommand{\ep}{\end{picture}}
\newcommand{\bfi}{\begin{figure}}
\newcommand{\efi}{\end{figure}}

\def\fun#1#2{\lower3.6pt\vbox{\baselineskip0pt\lineskip.9pt
\ialign{$\mathsurround=0pt#1\hfil##\hfil$\crcr#2\crcr\sim\crcr}}}

\begin{document}

\title{\LARGE \bf {Black Holes-Hedgehogs and Strings as Defects of the Universal Vacua}}
\author{\large \bf
  B.G. Sidharth ${}^{1}$\footnote{iiamisbgs@yahoo.co.in, birlasc@gmail.com},\,\,
  C.R. Das ${}^{2}$\footnote{das@theor.jinr.ru},\,\,
  L.V. Laperashvili ${}^{3}$\footnote{laper@itep.ru},\\
  \large \bf
  and
  H.B. Nielsen${}^{4}$\footnote{hbech@nbi.dk}\\\\
{\large \it ${}^{1}$ International Institute of Applicable Mathematics}\\
{\large \it and Information Sciences,}\\
{\large \it B.M. Birla Science Centre}\\
{\large \it Adarsh Nagar, 500063 Hyderabad, India}\\\\
{\large \it ${}^{2}$ Bogoliubov Laboratory of Theoretical Physics}\\
{\large \it Joint Institute for Nuclear Research}\\
{\large \it International Intergovernmental Organization,}\\
{\large \it Joliot-Curie 6, 141980 Dubna, Moscow region, Russia}\\\\
{\large \it ${}^{3}$ The Institute of Theoretical and
Experimental Physics,}\\
{\large\it National Research Center ``Kurchatov Institute'',}\\
{\large\it Bolshaya Cheremushkinskaya, 25, 117218 Moscow, Russia}\\\\
{\large \it ${}^{4}$ Niels Bohr Institute,}\\
{\large \it Blegdamsvej, 17-21, DK 2100 Copenhagen, Denmark}}

\date{}
\maketitle

\thispagestyle{empty}

\vspace{2cm}



\clearpage \newpage

\thispagestyle{empty}

\begin{abstract}
In the present paper, assuming the Multiple Point Principle (MPP)
as a new law of Nature, we considered the existence of the two
degenerate vacua of the Universe: a) the first Electroweak (EW)
vacuum at $v_1\approx 246$ GeV -- ``true vacuum", and b) the second Planck
scale ``false vacuum" at $v_2 \sim 10^{18}$ GeV. In these vacua we
investigated different topological defects. The main aim of this
paper is an investigation of the black-hole-hedgehogs
configurations as defects of the false vacuum. In the framework of
the $f(R)$ gravity, described by the Gravi-Weak unification model,
we considered a black-hole solution, which corresponds to a
``hedgehog" -- global monopole, that has been ``swallowed" by the
black-hole with mass core $M_{BH}\sim 10^{18}$ GeV and radius
$\delta\sim 10^{-21}$  GeV$^{-1}$. Considering the results
of the hedgehog lattice theory in the framework of the $SU(2)$
Yang-Mills gauge-invariant theory with hedgehogs in the Wilson
loops, we have used  the critical value of temperature for the
hedgehogs confinement phase ($T_c\sim 10^{18}$ GeV). This result
gave us the possibility to conclude that the SM  shows a new
physics (with contributions of the $SU(2)$-triplet Higgs bosons) at
the scale $\sim 10$ TeV. Theory predicts the stability of the
EW-vacuum and the accuracy of the MPP.
\end{abstract}

\clearpage \newpage

\section{Introduction}

The present article is devoted to studying of topological defects
of the universal vacua.

During the expansion after the Planck era, the early Universe
underwent a series of phase transitions as a result of which there
were arisen such vacuum topological defects (widely discussed in
literature) as monopoles or hedgehogs (point defects), strings
(line defects), bubbles and domain walls (sheet defects). These
topological defects appeared due to the breakdown of local or
global gauge symmetries.

This paper is essentially based on the discovery that a
cosmological constant of our Universe is extremely small, almost
zero \ct{1a,2a,3a}. We considered a Multiple Point Principle (MPP)
first suggested by D.L. Bennett and H.B. Nielsen \ct{1}, which
predicts the existence in Nature of several degenerate vacua with
very small energy density (cosmological constants).

The model developed in this article confirms the existence of the
two degenerate vacua of the Universe: The first (``true")
Electroweak (EW) vacuum with VEV $v_1\approx 246$ GeV, and the
second (``false") Planck scale vacuum with VEV $v_2\sim 10^{18}$
GeV.

The main idea of this paper is the investigation of hedgehog's
configurations \ct{1*,2*} as defects of the false vacuum. We have
shown that at superhigh (Planck scale) energies the
black-holes-hedgehogs are responsible for the creation of the
false vacuum of the Universe. In the framework of the $f(R)$
gravity, we have obtained a solution for a global monopole, which
is a black-hole-hedgehog at the Planck scale. Here we have used
the $f(R)$ gravity predicted by the Gravi-Weak unification model
previously developed by authors in papers
\ct{1gw,2gw,3gw,4gw,5gw}.

Using the results of Refs.\ct{BCK,C} obtained for the $SU(2)$
Yang-Mills theory of the gauge-invariant hedgehog-like structures
in the Wilson loops, we have considered the lattice theory giving
the critical value of temperature for the hedgehogs confinement
phase. Considering the hedgehog lattice theory, we have concluded
that hedgehogs can exist only at the energy scale $\mu \gtrsim
10^4$ GeV. Triplet Higgs fields $\Phi^a$ (with
$a=1,2,3$), which are responsible for the formation of hedgehogs,
can show a new physics at the scale $\sim 10$ TeV.

In Section 2 we reviewed the Multiple Point Principle (MPP)
suggested by D.L. Bennett and H.B. Nielsen \ct{1}.  In the
assumption of the existence of the two degenerate vacua
(Electroweak vacuum at $v_1 \approx 246$ GeV, and Planck scale one
at $v_2 \sim 10^{18}$ GeV), Froggatt and Nielsen \ct{FN} obtained
the first prediction of the top-quark and Higgs boson masses,
which was further improved by several authors in the next
approximations. Section 3 is devoted to the general properties of
topological defects of the universal vacua. We considered
topological defects in the ``false vacuum", which is presented as a
spherical bubble spontaneously produced in the de Sitter-like
universe. The space-time inside the bubble, which we refer to as a
``true vacuum", has the geometry of an open
Friedmann-Lemaitre-Robertson-Walker (FLRW) universe. Section 4 is
devoted to the Gravi-Weak unification (GWU) model
\ct{1gw,2gw,3gw,4gw,5gw} as an example of the $f(R)$ gravity.
Subsection 4.1 considers the existence of the de Sitter solutions
in the Planck phase. Subsection 4.2 is devoted to calculations of
parameters of the GWU-model, where we predicted the Planck scale
false vacuum VEV equal to $v_2\approx 6.28\times 10^{18}$ GeV. In
Section 5 we have investigated the hedgehog's configurations as
defects of the false vacuum. We obtained a solution for a
black-hole in the framework of the $f(R)$ gravity, which
corresponds to a global monopole ``swallowed" by a black-hole. The
metric around of the global monopole was considered in Subsection
5.1. The mass $M_{BH}$, radius $\delta$ and ``horizon radius" $r_h$
of the black-hole-hedgehog were estimated in Subsection 5.2.
Section 6 is devoted to the lattice-like structure of the false
vacuum which is described by a non-differentiable space-time: by a
foam of black-holes, having lattice-like structure, in which sites
are black-holes with ``hedgehog" monopoles inside them. This
manifold is described by a non-commutative geometry predicted an
almost zero cosmological constant. The phase transition from the
``false vacuum" to the ``true vacuum" was considered in Section 7,
where it was shown that the Electroweak spontaneous breakdown of
symmetry $SU(2)_L\times U(1)_Y \to U(1)_{el.mag}$ created new
topological defects of EW vacuum: the Abrikosov-Nielsen-Olesen
closed magnetic vortices (``ANO strings") of the Abelian Higgs
model and Sidharth's Compton phase objects. Then the ``true vacuum"
(EW-vacuum) again presents the non-differentiable manifold with
non-commutative geometry, and again has an almost zero
cosmological constant. Here we estimated the black-hole-hedgehog's
mass and radius: $M_{BH}\approx 3.65\times 10^{18}$ GeV and
$\delta\approx 0.29\lambda_{Pl}\approx 10^{-21}$ GeV$^{-1}$ near
the second vacuum $v_2$. In Subsection 7.1 we emphasize that due
to the energy conservation law, the vacuum density before the
phase transition is equal to the vacuum density after the phase
transition, and we have $$ \rho_{vac}({\rm at\,\, Planck\,\, scale}) =
\rho_{vac}({\rm at\,\, EW\,\, scale}).$$ Therefore, we confirmed the
Multiple Point Principle: we have two degenerate vacua $v_1$ and
$v_2$ with an almost zero vacuum energy (cosmological constants).
This means that our EW-vacuum, in which we live, is stable. The
Planck scale vacuum cannot be negative: $V_{eff}(min1) = V_{eff}(min2)$,
these potentials are equal exactly. In
Section 8 hedgehogs in Wilson loops of the $SU(2)$ Yang-Mills
theory, and phase transitions in this theory were investigated
using the results of Refs.~\ct{BCK,C}. Their lattice results gave
the critical value of the temperature for the hedgehog's
confinement phase: $\beta_{crit} \approx 2.5$, and this result
gives the value of critical temperature $T_c\sim 10^{18}$ GeV. In
Section 9 we show that the hedgehog's confinement happens at
energy $\sim 10$ TeV, which is a threshold energy of the
production of a pair of the $SU(2)$-triplet Higgs bosons. In Section
10 we reviewed the problem of the vacuum stability in the Standard
Model. In Section 11 we show that hedgehogs can contribute at
energy scale $\mu > 10^4$ GeV. Therefore, a triplet Higgs field
$\Phi^a$ provides a new physics at the scale $\sim 10$ TeV. In
this Section 11 we predict an exact stability of the EW-vacuum and
the accuracy of the MPP.

\section{Degenerate vacua of the Universe}

This paper is based on the new law of Nature named {\un{the
Multiple Point Principle (MPP)}} which was suggested by D.L.
Bennett and H.B. Nielsen in Ref.\ct{1}. The MPP means: {\it There
exist in Nature several degenerate vacua with very small energy
density, or cosmological constants.}

Vacuum energy density of our Universe is the Dark Energy (DE). It
is related with cosmological constant $\Lambda$ by the following
way:
\be
\rho_{DE} = \rho_{vac} = (M^{red}_{Pl})^2\Lambda,  \lb{1} \ee
where $M^{red}_{Pl}$ is the reduced Planck mass: $M^{red}_{Pl}\simeq 2.43 \times 10^{18}$ GeV.
At present, cosmological measurements give:
\be  \rho_{DE} \simeq (2\times 10^{-3}\,\, {\rm{ eV}})^4,  \lb{2} \ee
which means a tiny value of the cosmological constant:
\be \Lambda\simeq 10^{-84}\,\, {\rm GeV}^2. \lb{3} \ee
This tiny value of $\rho_{DE}$ was first predicted by B.G.
Sidharth in 1997 year \ct{1a,2a}. In the 1998 year S. Perlmutter, B.
Schmidt and A. Riess \ct{3a} were awarded the Nobel Prize for
the discovery of the accelerating expansion of the Universe.

Having an extremely small cosmological constant of our Universe,
Bennett, Froggatt and Nielsen \ct{1,FN,2,3} assumed to consider
only zero, or almost zero, cosmological constants for all vacua
existing in Nature.

The MPP theory was developed in a lot of papers by H.B. Nielsen
and his collaborators (see for example, Refs.
\ct{1,FN,2,3,4,5,6,7,8,9,10,11,12,12a,12b,13,14,15} and recent
Refs. \ct{18,19,20,21} by other authors).

Restricted to pure Standard Model (SM) we assumed the existence of
only three vacua:

\begin{enumerate}
\item {\bf Present Electroweak vacuum}, ``true vacuum", in which we live.\\
It has vacuum expectation value (VEV) of the Higgs field equal to:
\be v_1 = \langle \phi_H\rangle  \approx 246\,\, {\rm{GeV}}. \lb{4}
\ee
\item {\bf High Higgs field vacuum}, ``false vacuum" -- Planck
scale vacuum, which has the following VEV:
\be v_2 = \langle \phi_H\rangle  \sim 10^{18}\,\, {\rm{GeV}}. \lb{5}
\ee
\item {\bf Condensate vacuum.} This third vacuum is a very speculative possible state
inside the pure SM, which contains a lot of strongly bound states, each
bound from 6 top + 6 anti-top quarks (see Refs.~\ct{25,26,27,28,29}).
\end{enumerate}

From experimental results for these three vacua, cosmological
constants -- minima of the Higgs effective potentials $V_{eff} (\phi_H)$ -- are
not exactly equal to zero. Nevertheless, they are extremely small.
By this reason, Bennett, Froggatt and Nielsen assumed to consider zero cosmological
constants as a good approximation. Then according to the MPP, we have a model of pure SM
being fine-tuned in such a way that these three vacua proposed have just zero energy density.

If the effective potential has three degenerate minima, then the
following requirements are satisfied:
\be
V_{eff} (\phi^2_{min1}) =  V_{eff} (\phi^2_{min2}) =  V_{eff} (\phi^2_{min3}) = 0,  \lb{6}
\ee
and
\be
V'_{eff} (\phi^2_{min1}) =  V'_{eff} (\phi^2_{min2}) =  V'_{eff} (\phi^2_{min3}) = 0, \lb{7}
\ee
where
\be
V'(\phi^2) = \frac{\partial V}{\partial \phi^2}. \lb{8}
\ee
Here we assume that:
\be
V_{eff} (\phi^2_{min1}) = V_{present}, \quad V_{eff} (\phi^2_{min2}) = V_{high\,\, field}, \quad
{\rm{and}} \quad V_{eff} (\phi^2_{min3}) = V_{condensate}.     \lb{9}
\ee
Assuming the existence of the two degenerate vacua in the SM:

a) the first Electroweak vacuum at $v_1 \approx 246$ GeV, and

b) the second Planck scale vacuum at $v_2 \sim 10^{18}$ GeV,\\
Froggatt and Nielsen predicted in Ref.~\ct{FN} the top-quark and
Higgs boson masses:
\be M_t = 173 \pm 5\,\, {\rm { GeV}}; \quad M_H = 135 \pm 10\,\, {\rm
{ GeV}}. \lb{10} \ee

In Fig.~1 it is shown the existence of the second (non-standard)
minimum of the effective Higgs potential in the pure SM at the Planck
scale.

\section{Topological defects of the universal vacua}

Topological structures in fields are as important as the fields
themselves. The presence of defects determines special features of
the vacuum.

It is well known that in the early Universe topological defects
may be created in the vacuum during the vacuum phase transitions.
The early Universe underwent a series of phase transitions, each
one spontaneously breaking some symmetry in particle physics and
giving rise to topological defects of some kind, which can play an
essential role throughout the subsequent evolution of the
Universe.

In the context of the General Relativity, Barriola and Vilenkin (see Ref.~\ct{31})
studied the gravitational effects of a global monopole as a spherically
symmetric topological defect. The authors found, that the gravitational effect of 
the global monopole is repulsive in nature. Thus, one may expect that the global monopole
and cosmological constants are connected through their common manifestation as the
origin of repulsive gravity. Moreover, both cosmological constant and vacuum expectation
value (VEV) are connected while the VEV is connected to the topological defects.
All these points lead us to a simple conjecture: There must be a common connection
among the cosmological constant, topological defects and the vacuum expectation
values (VEVs).

Different phase transitions have resulted during the expansion of
the early Universe after the Planck era. They produced the
formation of the various kind of topological defects: {\bf  point
defects} (monopoles, hedgehogs, etc.); {\bf line defects}
(strings, vortices), and {\bf sheet defects } (for example, domain
walls). The topology of the vacuum manifold dictates the nature of
these topological defects, appearing due to the breakdown of local
or global gauge symmetries.

In the present paper, we shall discuss another potentially observable
manifestation of topological defects. It has been shown in Ref.~\ct{32}
that topological defects, like spherical domain walls and circular loops
of cosmic string, can be spontaneously produced in a de Sitter-like
universe. The initial radii of walls and strings are close to the de Sitter
horizon, which corresponds to the Universe radius:
\be R_{un}\simeq R_{de\,\, Sitter\,\, horizon}\simeq 10^{28}\,\, {\rm{cm}}.  \lb{11}
\ee
In the present paper we study the evolution of the two bubbles:
one having a ``false vacuum", and the other one having a ``true
vacuum". The bubble, which we shall refer to as the {\bf false
vacuum}, to be a de Sitter space with a constant expansion rate
$H_F$ . It is convenient to use flat de Sitter coordinates to
describe the background of the inflating false vacuum:
\be
ds^2 = dt^2 - e^{2H_F t} (dr^2 + r^2d\Omega^2),  \lb{12}
\ee
where
\be   d\Omega^2 = d\theta^2 + \sin^2\theta d\phi^2.   \lb{13}
\ee
The space-time inside the bubble, which we shall refer to as a
{\bf true vacuum}, has the geometry of an open
Friedmann-Lemaitre-Robertson-Walker (FLRW) universe (see for
example review \ct{Cop}):
 \be ds^2 = d\tau^2 - a(\tau)^2(d\xi^2 + \sinh^2\xi d\Omega^2),
\lb{14} \ee
where $a(\tau)$ is a scale factor with cosmic time $\tau$. In the
true vacuum we have a constant expansion rate $H_T$, which has the
meaning of the slow-roll inflation rate inside the bubble at early
stage of its evolution.

Cosmological theory of bubbles was developed in a lot of papers by
A. Vilenkin and his collaborators (see for example
Refs.\ct{32,33,34}). The physical properties of defects depend on
the embedding vacuum.

\section{Gravi-Weak unification and hedgehogs as defects of the
false vacuum}

In the paper \ct{1gw} (using the ideas of Refs.\ct{34a} and
\ct{34b}) we have considered a $Spin(4,4)$-group of the gravi-weak
unification which is spontaneously broken into the
$SL(2,C)^{(grav)}\times SU(2)^{(weak)}$. Such a model was
constructed in agreement with experimental and astrophysical
results. We assumed that after the Bing Bang there existed a
Theory of the Everything (TOE) which rapidly was broken down to
the direct product of the following gauge groups:

\bea G_{(TOE)} &\to& G_{(GW)}\times U(4) \to SL(2,C)^{(grav)}\times SU(2)^{(weak)}\times U(4)\nonumber\\
&\to& SL(2,C)^{(grav)}\times SU(2)^{(weak)}\times SU(4)\times U(1)_Y\nonumber\\
&\to& SL(2,C)^{(grav)}\times SU(2)^{(weak)}\times SU(3)_c\times U(1)_{(B-L)}\times U(1)_Y\nonumber\\
&\to& SL(2,C)^{(grav)}\times SU(3)_c\times SU(2)_L \times U(1)_Y\times U(1)_{(B-L)}\nonumber\\
&\to& SL(2,C)^{(grav)}\times G_{SM} \times U(1)_{(B-L)}.\nonumber\eea
And below the
see-saw scale ($M_R\sim 10^9\sim 10^{14}$ GeV) we have
the SM group of symmetry:
$$ G_{SM} = SU(3)_c\times SU(2)_L \times U(1)_Y.$$
The action $S_{(GW)}$ of the Gravi-Weak unification (obtained in
Ref.~\ct{1gw}) was given by the following expression:
\bea S_{(GW)} &=& - \frac 1{g_{uni}}\int_{\mathfrak
M}d^4x\sqrt{-g}\left[\frac 1{16}\left(R|\Phi|^2 - \frac 32|\Phi|^4\right)\right.\nonumber\\
&&+\left. \frac 1{16}\left(aR_{\mu\nu}R^{\mu\nu} + bR^2\right) + \frac 12{\cal
D}_{\mu}\Phi^\dag{\cal D}^{\mu}\Phi + \frac 14
F_{\mu\nu}^iF^{i\,\mu\nu}\right],  \lb{15} \eea
where $g_{uni}$ is a parameter of the graviweak unification,
parameters $a,b$ (with $a+b=1$) are ``bare" coupling constants of
the higher derivative gravity, $R$ is the Riemann curvature
scalar, $R_{\mu\nu}$ is the Ricci tensor, $|\Phi|^2 =
\Phi^a\Phi^a$ is a squared triplet Higgs field, where $\Phi^a$
(with $a=1,2,3$) is an isovector scalar belonging to the adjoint
representation of the $SU(2)$ gauge group of symmetry. In
Eq.(\ref{15}):
\be   {\cal D}_{\mu}\Phi^a = \partial_{\mu}\Phi^a +
g_2\epsilon^{abc}A_{\mu}^b\Phi^c       \lb{16} \ee
is a covariant derivative, and
\be   F_{\mu\nu}^a = \partial_{\mu}A_{\nu}^a -
\partial_{\nu}A_{\mu}^a + g_2\epsilon^{abc}A_{\mu}^bA_{\nu}^c  \lb{17} \ee
is a curvature of the gauge field $A_{\mu}^a$ of the $SU(2)$
Yang-Mills theory. The coupling constant $g_2$ is a ``bare"
coupling constant of the $SU(2)$ weak interaction.

The GW action (\ref{15}) is a special case of the $f(R)$ gravity
\ct{34c,34d,34e}  when:
\be f(R) = R|\Phi|^2.    \lb{18} \ee
In a general case of the $f(R)$ gravity, the action contains
matter fields and can be presented by the following expression:
\be S = \frac{1}{2\kappa}\int d^4x \sqrt{-g}\,f(R) + S_{grav} +
S_{gauge} + S_m, \lb{19} \ee
where $S_m$ corresponds to the part of the action associated with
matter fields, fermions and Higgs fields.

Using the metric formalism, we obtain the following field
equations:
\be F(R)R_{\mu\nu} - \frac 12 f(R)g_{\mu\nu} - \nabla_{\mu}
\nabla_{\nu}F(R) + g_{\mu\nu}\Box F(R) = \kappa T_{\mu\nu}^m,
                             \lb{20} \ee
where:
\be F(R)\equiv \frac{df(r)}{dr},    \lb{21} \ee
$\kappa = 8\pi G_N$, $G_N$ is the gravitational constant, and
$T^m$ is the energy-momentum tensor derived from the matter action
$S_m$.

\subsection{The existence of the de Sitter solutions at the early time of
the Universe}

It is well-known that at the early time of the Universe an
acceleration era is described by the de Sitter solutions (see for
example \ct{34f,34g}). The investigation of the problem that de
Sitter solutions exist in the case of the action (15) was
considered by authors of Ref.\ct{34a}. Our model \ct{1gw} is a
special case of the more general $SU(N)$ model \ct{34a}, and we can
assume that the Universe is inherently de Sitter. Then the
4-spacetime is a hyperboloid in a 5-dimensional Minkowski space
under the constraint:
\be x^2_0 + x^2_1 + x^2_2 + x^2_3 + x^2_4 = r^2_{dS}, \lb{22} \ee
where $r_{dS}$ is a radius of a curvature of the de Sitter space,
or simply ``the de Sitter radius". The Hubble expansion of the
Universe is then viewed as a process that approaches the
asymptotic limit of a pure space which is de Sitter in nature,
evidenced that the cosmological constant $\Lambda$ describes the
dark energy (DE) substance, which has become dominant in the
Universe at later times:
\be \Omega_{DE} = \frac{\rho_{DE}}{\rho_{crit}}\simeq 0.75,
                                   \lb{23} \ee
where $\rho_{DE}$ is the dark energy density and the critical
density is:
\be \rho_{crit} = \frac{3H_0^2}{8\pi G_N}\simeq 1.88\times
10^{-29\,\, }H_0^2,
                                   \lb{24} \ee
where $H_0$ is the Hubble constant:
\be H_0 \simeq 1.5 \times 10^{-42}\,\, {\rm{GeV}}.
                                   \lb{25} \ee
Identifying the Einstein tensor as
\be   G_{\mu\nu} = - \frac{3}{r_{dS}^2} g_{\mu\nu},
                                   \lb{26} \ee
we see that the only nontrivial component that satisfies this
equation is a constant for the Ricci scalar:
\be   R_0 = \frac{12}{r_{dS}^2}.
                                   \lb{27} \ee
As it was shown in Ref.\ct{34a}, the nontrivial vacuum solution to
the action (15) is de Sitter spacetime with a non-vanishing Higgs
vacuum expectation value (VEV) of the triplet Higgs scalar field
$\Phi$: $v_2 = \langle\Phi\rangle = \Phi_0$. The standard Higgs potential in
Eq.(15) has an extremum at $\Phi_0 = R/3$ (with $R > 0$),
corresponding to a de Sitter spacetime background solution:
\be   R = R_0 =\frac{12}{r_{dS}^2} = 3v_2^2,
                                   \lb{28} \ee
which implies vanishing curvature:
\be F_0 = \frac 12 R_0 - \frac 1{16}\Sigma_0\Phi_0^2   \lb{29} \ee
solving the field equations $DF = dF + [A, F] = 0$, and strictly
minimizing the action (15).

Based on this picture, the origin of the cosmological constant
(and DE) is associated with the inherent spacetime geometry, and
not with vacuum energy of particles (we consider their
contributions later). We note that as a fundamental constant under
the de Sitter symmetry, $r_{dS}$ is not a subject to quantum
corrections. Local dynamics exist as fluctuations with respect to
this cosmological background. In general, the de Sitter space may
be inherently unstable. The quantum instability of the de Sitter
space was investigated by various authors. Abbott and Deser
\ct{34k} have shown that de Sitter space is stable under a
restricted class of classical gravitational perturbations. So any
instability of the de Sitter space may likely have a quantum
origin. Ref.\ct{34l} demonstrated through the expectation value of
the energy-momentum tensor for a system with a quantum field in a
de Sitter background space that in general, it contains a term that
is proportional to the metric tensor and grows in time. As a
result, the curvature of the spacetime would decrease and the de
Sitter space tends to decay into the flat space (see
Ref.\ct{34m}). The decay time of this process  is of the order of
the de Sitter radius:
\be   \tau \sim r_{dS}\simeq 1.33\,\, H_0^{-1}.  \lb{31} \ee
Since the age of our universe is smaller than $r_{dS}$, we are
still observing the accelerating expansion in action.

Of course, we can consider the perturbation solutions of the de
Sitter solution but these perturbations are very small
\ct{34f,34g}.

\subsection{Parameters of the Gravi-Weak unification model}

Assuming that at the first stage of the evolution (before the
inflation), the Universe had the de Sitter spacetime -- maximally
symmetric Lorentzian manifold with a constant and positive
background scalar curvature $R$ -- we have obtained the following
relations from the action (\ref{15}):

1) The vacuum expectation value $v_2$ -- the VEV of ``the false
vacuum" -- is given by the de Sitter scalar curvature $R$:
\be     v_2^2 = \frac{R}{3}.  \lb{33} \ee

2) At the Planck scale the squared coupling constant of the weak
interaction is:
\be     g_2^2 = g_{uni}.  \lb{34} \ee
The replacement:
\be   \frac{\Phi^a}{g_2} \to \Phi^a  \lb{35} \ee
leads to the following GW-action:
\be  S_{(GW)} = - \int_{\mathfrak M}d^4x\sqrt{-g}\left(\frac
{R}{16}|\Phi|^2 - \frac{3g_2^2}{32}|\Phi|^4 + \frac 12{\cal
D}_{\mu}\Phi^\dag{\cal D}^{\mu}\Phi + \frac {1}{4g_2^2}
F_{\mu\nu}^iF^{i\,\mu\nu}  + {\rm{grav.\,\, terms}}\right),
\lb{15x} \ee
Now considering the VEV of the false vacuum as $v=v_2$, we have:
\be v^2 = \frac{R}{3g_2^2}.  \lb{33a} \ee
The Einstein-Hilbert action of general relativity with the
Einstein's cosmological constant $\Lambda_E$ is given by the
following expression:
\be   S_{EH} = - \frac 1{\kappa}\int d^4x \sqrt{-g}\left(\frac {R}{2} -
\Lambda_E\right).   \lb{36} \ee

3) The comparison of the Lagrangian $L_{EH}$ with the Lagrangian
given by Eq.(\ref{15x}) near the false vacuum $v$ leads to the
following relations for the Newton's gravitational constant $G_N$
and reduced Planck mass:
\be     (M_{Pl}^{red})^2 = (8\pi G_N)^{-1} = \frac{1}{\kappa} =
\frac{v^2}{8}.    \lb{37} \ee

4) Then we have:
\be   v = 2\sqrt 2 M_{Pl}^{red}\approx 6.28\times 10^{18}\,\,
{\rm{GeV}},         \lb{38} \ee
and
\be  \Lambda_E = \frac{3g_2^2}{4} v^2.  \lb{39} \ee
Eq.(\ref{37}) gives:
\be \frac{1}{\kappa}\Lambda_E = \frac{3g_2^2}{32}v^4.  \lb{39}
\ee
Using the well-known in literature renormalization group equation
(RGE) for the SU(2) running constant $\alpha_2^{-1}(\mu)$, where
$\alpha_2 = g_2^2/4\pi$ and $\mu$ is the energy scale, we can use
the extrapolation of this value to the Planck scale \ct{6,7} and
obtain the following result:
\be  \alpha_2(M_{Pl}) \sim \frac{1}{50},\quad g_{uni} = g_2^2 =
4\pi \alpha_2(M_{Pl}) \approx 4\pi \times 0.02 \approx 0.25.
\lb{40} \ee

\section{The solution for the black-holes-hedgehogs}

A global monopole is described by the part $L_h$ of the Lagrangian
$L_{(GW)}$ given by the action (\ref{15x}), which contains the
$SU(2)$-triplet Higgs field $\Phi^a$, VEV of the second vacuum
$v_2=v$ and cosmological constant $\Lambda=\Lambda_E$:
\bea L_h &=&  - \frac{R}{16}|\Phi|^2 + \frac{3g_2^2}{32}|\Phi|^4 - \frac
12\partial_{\mu}\Phi^a\partial^{\mu}\Phi^a + \Lambda_E\nonumber\\
&=& -
\frac 12
\partial_{\mu}\Phi^a\partial^{\mu}\Phi^a +
\frac{\lambda}{4}\left(|\Phi|^2 - v^2\right)^2 + \frac{\Lambda_E}{\kappa} -
\frac{\lambda}{4}v^4\nonumber\\
&=& - \frac 12
\partial_{\mu}\Phi^a\partial^{\mu}\Phi^a +
\frac{\lambda}{4}\left(|\Phi|^2 - v^2\right)^2. \lb{41} \eea
Here  we have:
\be     \lambda = \frac {3g_2^2}{8}.   \lb{42} \ee
Substituting in Eq.(\ref{42}) the value $g_2^2\approx 0.25$ given
by Eq.(\ref{40}), we obtain:
\be     \lambda \approx \frac{3}{32}.   \lb{43} \ee
Eq.(\ref{39}) gives:
\be \frac{\Lambda_E}{\kappa} = \frac{3g_2^2}{32}v^4 =
\frac{\lambda}{4}v^4,    \lb{44} \ee
and in Eq.(\ref{41}) we have the compensation of the Einstein's
cosmological term. Then
\be    L_h =  - \frac 12
\partial_{\mu}\Phi^a\partial^{\mu}\Phi^a + V(\Phi),
                                 \lb{46} \ee
where the Higgs potential is:
\be  V(\Phi) = \frac{\lambda}{4}\left(|\Phi|^2 - v^2\right)^2. \lb{47} \ee
This potential has a minimum at $\langle|\Phi|\rangle_{min}=v$, in which it
vanishes:
\be V\left(|\Phi|^2_{min}\right) = V'\left(|\Phi|^2_{min}\right) = 0, \lb{48} \ee
in agreement with the MPP conditions (\ref{6}) and (\ref{7}).

The field configurations describing a monopole-hedgehog
\ct{1*,2*} are:
\bea \Phi^a &=& v w(r )\frac{x^a}{r},  \lb{1m}  \nonumber\\
 A_{\mu}^a &=& a(r )\epsilon_{\mu ab}\frac{x^b}{r},  \lb{2m} \eea
where $x^ax^a = r^2$ with $(a = 1, 2, 3)$, $w(r)$ and $a(r)$ are
some structural functions. This solution is pointing radially.
Here $\Phi^a$ is parallel to $\hat{r}$ -- the unit vector in the
radial, and we have a ``hedgehog" solution of Refs.~\ct{1*,2*}. The
terminology ``hedgehog" was first suggested by Alexander Polyakov
in Ref.~\ct{2*}. 

The field equations for $\Phi^a$ in the flat metric reduces to a
single equation for $w(r)$:
\be  w'' + \frac 2{r}w' - \frac 2{r^2}w - \frac{w(w^2
-1)}{\delta^2} = 0,     \lb{3m} \ee
where $\delta$ is the core radius of the hedgehog. In the flat
space the hedgehog's core has the following size:
\be  \delta \sim \frac{1}{\sqrt \lambda v}.     \lb{4m} \ee
The function $w(r)$ grows linearly when $r < \delta$ and
exponentially approaches unity as soon as $r > \delta$ . Barriola
and Vilenkin \ct{31} took $w = 1$ outside the core which is an
approximation to the exact solution. As a result, the functions
$w(r)$ and $a(r)$ are constrained by the following conditions:
\bea  w(0) &=& 0, \quad {\rm{and}} \quad w(r)\to 1 \quad
{\rm{when}}\quad r\to \infty,    \lb{5m} \nonumber\\
 a(0) &=& 0, \quad {\rm{and}} \quad a(r)\sim - \frac 1{r} \quad
{\rm{when}}\quad r\to \infty.    \lb{6m} \eea

\subsection{The metric around of the global monopole}

The most general static metric around of the global monopole is a
metric with spherical symmetry:
\be ds^2 = B(r)dt^2 - A(r)dr^2 -r^2(d\theta^2 + sin^2\theta
d\varphi^2).     \lb{7m} \ee
For this metric the Ricci tensor has the following non-vanishing
components:
\bea R_{tt} &=& -\frac{B''}{2A} + \frac{B'}{4A}\left(\frac{A'}{A} +
\frac{B'}{B}\right) - \frac{1}{r}\frac{B'}{A},\nonumber\\
  R_{rr} &=& \frac{B''}{2B} + \frac{B'}{4B}\left(\frac{A'}{A} +
\frac{B'}{B}\right) - \frac{1}{r}\frac{A'}{A},\nonumber\\
  R_{\theta\theta} &=& - 1 + \frac{r}{2A}\left( - \frac{A'}{A} +
\frac{B'}{B}\right) + \frac{1}{A},\nonumber\\ R_{\varphi\varphi} &=&
\sin^2\theta R_{\theta\theta}.   \lb{8m} \eea
The energy-momentum tensor of the monopole is given by
\bea  T^t_t &=& v^2\frac{w'^2}{2A} + v^2\frac{w^2}{r^2} + \frac 14
\lambda v^4 (w^2 - 1)^2, \nonumber\\
  T^r_r &=& - v^2\frac{w'^2}{2A} + v^2\frac{w^2}{r^2} + \frac 14
\lambda v^4 (w^2 - 1)^2,  \nonumber\\ T^\theta_\theta =
T^\varphi_\varphi &=& v^2\frac{w'^2}{2A} + \frac 14 \lambda v^4 (w^2
- 1)^2.       \lb{9m} \eea
Here $\kappa=1$.

Considering the approximation used by Barriola and Vilenkin in
Ref.\ct{31}, we obtain an approximate solution for
monopole-hedgehog taking $w=1$ out the core of the hedgehog (see
also Refs.\ct{35,36,37,ShiLi,Car}). In this case scalar curvature
$R$ is constant and Eq.(\ref{20}) comes down to the Einstein's
equation:
\bea  \frac{1}{A}\left(\frac{1}{r^2} - \frac{1}{r}\frac{A'}{A}\right) -
\frac{1}{r^2} &=& \frac{1}{v^2}T^t_t,   \lb{10m} \\
  \frac{1}{A}\left(\frac{1}{r^2} + \frac{1}{r}\frac{B'}{B}\right) -
\frac{1}{r^2} &=& \frac{1}{v^2}T^r_r,   \lb{11m} \eea
where the energy-momentum tensor is given by the following
approximation:
\bea T^t_t &=& T^r_r \approx \frac{v^2}{r^2}, \nonumber\\
 T^\theta_\theta &=& T^\varphi_\varphi = 0. \lb{12mx} \eea
Taking into account Eq.(\ref{12mx}), we obtain the following result
by substraction of Eqs.(\ref{10m}) and (\ref{11m}):
\be    \frac{A'}{A} + \frac{B'}{B} = 0, \lb{12m} \ee
and then asymptotically (when $r\to \infty$) we have:
\be    A\approx B^{-1}. \lb{13m} \ee
From Eq.(\ref{10m}) we obtain a general relation for the function
$A(r)$:
\be   A^{-1}(r) = 1 - \frac{1}{r}\int_0^r T^t_t\, r^2dr. \lb{14m}
\ee
In the limit $r\to \infty$ we obtain:
\bea   A(r) &=& 1 - \kappa v^2 - \frac{2G_N\,M}{r} +... \lb{15m} \nonumber\\
   B(r) &=& \left(1 - \kappa v^2 - \frac{2G_N\,M}{r} +...\right)^{-1}
                                        \lb{16m} \eea

\subsection{The mass, radius and horizon radius of the black-hole-hedgehog}

Eq.(\ref{14m}) suggests the following equation for the hedgehog
mass $M$:
\be    M = 8\pi \int_0^{\infty} T^t_t\, r^2dr,
                                    \lb{17m} \ee
or
\be    M = 8\pi v^2\int_0^{\infty} \left(w'^2 + \frac{w^2 - 1}{r^2} +
\frac{(w^2 - 1)^2}{4\delta^2}\right)\, r^2dr.
                                    \lb{18m} \ee
The function $w(r)$ was estimated in Ref.\ct{ShiLi} at $r <
\delta$:
\be    w(r)\approx 1 - \exp\left(-\frac{r}{\delta}\right),
                                    \lb{19m} \ee
and we obtain an approximate value of the hedgehog mass:
\be  M = M_{BH} \approx - 8\pi v^2 \delta.
                                    \lb{20m} \ee
There is a repulsive gravitational potential due to this negative
mass. A freely moving particle near the core of the black-hole
experiences an outward proper acceleration:
\be   \ddot{r} = - \frac{G_NM}{r} = \frac{G_N|M|}{r}.
                                    \lb{21m} \ee
We have obtained a global monopole with a huge mass (\ref{20m}),
which has a property of the hedgehog. This is a black-hole
solution, which corresponds to a global monopole-hedgehog that has
been ``swallowed" by a black-hole. Indeed, we have obtained the
metric result by M. Barriola et.al. \ct{31} like:
\be ds^2 = \left(1 - \kappa v^2 + \frac{2G_N\,|M|}{r}\right)dt^2 -
\frac{dr^2}{1 - \kappa v^2 + \frac{2G_N\,|M|}{r}} - r^2\left(d\theta^2
+ \sin^2\theta d\varphi^2\right).     \lb{22m} \ee
A black hole has a horizon. A horizon radius $r_h$ is found by
solving the equation $A(r_h) = 0$:
\be  1 - \kappa v^2 + \frac{2G_N\,|M|}{r_h} = 0,  \lb{23m} \ee
and we have a solution:
\be  r_h =  \frac{2 G_N\,|M|}{\kappa v^2 - 1}.      \lb{24m} \ee
According to Eq.~(\ref{37}), $\kappa v^2 = 8$, and we obtain the
black-hole-hedgehog with a horizon radius:
\be  r_h =  \frac 27 G_N\,|M| = \frac 27\times
\frac{\kappa}{8\pi}\times|M| = \frac
27\times\frac{\kappa}{8\pi}\times 8\pi v^2\delta \approx
\frac{16}{7}\delta \approx 2.29\delta. \lb{25m} \ee
We see that the horizon radius $r_h$ is more than the hedgehog
radius $\delta$:
$$r_h > \delta,$$
and our concept that ``a black hole contains the hedgehog" is
justified.

\section{Lattice-like structure of the false vacuum and non-commutativity}

Now we see, that at the Planck scale the false vacuum of the
Universe is described by a non-differentiable space-time: by a
foam of black-holes, having lattice-like structure, in which sites
are black-holes with ``hedgehog" monopoles inside them. This
manifold is described by a non-commutative geometry
\ct{1a,2a,38,42,43,44,45,46,47}.

In Refs.~\ct{1a,2a} B.G. Sidharth predicted:
\begin{enumerate}
\item[1.] That a cosmological constant is given by a tiny value:
\be  \Lambda \sim H_0^2,   \lb{15a} \ee
where $H_0$ is the Hubble rate in the early Universe:
\be  H_0 \simeq 1.5 \times 10^{-42}\,\, {\rm GeV}.  \lb{16a} \ee
\item[2.] That a Dark Energy density is very small:
\be  \rho_{DE} \simeq 10^{-12}\,\, {\rm eV}^4 = 10^{-48}\,\, {\rm GeV}^4;  \lb{17a} \ee
\item[3.] That a very small DE-density provides an accelerating
expansion of our Universe after the Big Bang.
\end{enumerate}

Sidharth proceeded from the following points of view \ct{44} :
Modern Quantum Gravity \ct{Rov} (Loop Quantum Gravity, etc.,) deal
with a non-differentiable space-time manifold. In such an
approach, there exists a minimal space-time cut off
$\lambda_{min}$, which leads to the non-commutative geometry, a
feature shared by the Fuzzy Space-Time also.

If the space-time is fuzzy, non-differentiable, then it has to be
described by a non-commutative geometry with the coordinates
obeying the following commutation relations:
\be  [dx^{\mu}, dx^{\nu}] \approx \beta^{\mu\nu}l^2 \neq 0. \lb{20a} \ee
Eq.~(\ref{20a}) is true for any minimal cut off $l$.

Previously the following commutation relation was considered by H.S. Snyder \ct{snyd}:
\be   [x, p] = \hslash \left( 1 + \left(\frac{l}{\hslash}
\right)^2p^2\right),\,\, etc.,  \lb{21a} \ee
which shows that effectively 4-momentum $p$ is replaced by
\be   p \to  p\left( 1 + \left(\frac{l}{\hslash}\right)^2p^2\right)^{-1}.
                           \lb{22a} \ee
Then the energy-momentum formula becomes as:
\be E^2 = m^2 + p^2\left( 1 + \left(\frac{l}{\hslash}\right)^2p^2\right)^{-2},
                                         \lb{23a} \ee
or
\be E^2 \approx m^2 + p^2 - 2\left(\frac{l}{\hslash}\right)^2p^4. \lb{24a} \ee
In such a theory the usual energy momentum dispersion relations are modified \ct{45,46}.
In the above equations, $l$ stands for a minimal (fundamental) length, which could
be the Planck length $\lambda_{Pl}$, or for more generally -- Compton wavelength
$\lambda_c$.

Writing Eq.~(\ref{24a}) as
\be    E = E' + E'',   \lb{25a} \ee
where $E'$ is the usual (old) expression for energy, and $E''$ is
the new additional term in modification. $E''$ can be easily
verified as $E'' = - m_bc^2$ -- for boson fields, and $E'' =
+m_fc^2$ -- for fermion fields with masses $m_b,m_f$,
respectively. These formulas help to identify the DE density, what
was first realized by B.G. Sidharth in Ref.~\ct{2a}.

DE density is a density of the quantum vacuum energy of the
Universe. Quantum vacuum, described by Zero Point Fields (ZPF)
contributions, is the lowest state of any Quantum Field Theory
(QFT), and due to the Heisenberg's principle has an infinite
value, which is renormalizable.

As it was pointed out in Refs.~\ct{Zel,42}, the quantum vacuum of
the Universe can be a source of the cosmic repulsion. However, a
difficulty in this approach has been that the value of the
cosmological constant turns out to be huge \ct{Zel}, far beyond
the value which is observed by astrophysical measurements. This
phenomenon has been called ``the cosmological constant problem"
\ct{Wein}.

A global monopole is a heavy object formed as a result of the
gauge-symmetry breaking during the phase transition of the
isoscalar triplet $\Phi^a$ system. The black-holes-hedgehogs are
similar to elementary particles, because a major part of their
energy is concentrated in a small region near the monopole core.
Assuming that the Planck scale false vacuum is described by a non-
differentiable space-time having lattice-like structure, where
sites of the lattice are black-holes with ``hedgehog" monopoles
inside them, we describe this manifold by a non-commutative
geometry with a minimal length $l=\lambda_{Pl}$. Using the
non-commutative theory of the discrete space-time, B.G. Sidharth
predicted in Refs.~\ct{2a,42} a tiny value of the cosmological
constant: $\Lambda \simeq 10^{-84}$ GeV$^2$ as a result of
the compensation of ZPF contributions by non-commutative
contributions of the hedgehog lattice.

\section{The phase transition from the ``false vacuum" to the ``true vacuum"}

In the Guendelman-Rabinowitz theory \ct{35} of the universal
vacua, the authors investigated the evolution of the two phases:

\begin{enumerate}
\item one being the ``false vacuum" (Planck scale vacuum), and
\item the other -- the ``true vacuum" (EW-scale vacuum).
\end{enumerate}

By cosmological theory, the Universe exists in the Planck scale
phase for extremely short time. By this reason, the Planck scale
phase was called ``the false vacuum". The presence of hedgehogs
as defects is responsible for the destabilization of the
false vacuum. The decay of the false vacuum is accompanied by the
decay of the black-holes-hedgehogs. These configurations are
unstable, and at some finite cosmic temperature which is called
the critical temperature $T_c$, a system exhibits a spontaneous
symmetry breaking, and we observe a phase transition from the
bubble with the false vacuum to the bubble with the true vacuum.
After the phase transition, the Universe begins its evolution
toward the low energy Electroweak (EW) phase. Here the Universe
underwent the inflation, which led to the phase having the VEV
$v_1\approx 246$ GeV. This is a ``true" vacuum, in which we live.

Guendelman and Rabinowitz \ct{35} also
allowed a possibility to consider an arbitrary domain wall between
these two phases. During the inflation, domain wall annihilates,
producing gravitational waves and a lot of the SM particles,
having masses.

The Electroweak spontaneous breakdown of symmetry $SU(2)_L\times
U(1)_Y \to U(1)_{el.mag}$ leads to the creation of the topological
defects in the EW vacuum. They are the Abrikosov-Nielsen-Olesen
closed magnetic vortices (``ANO strings") of the Abelian Higgs
model \ct{48,49}, and Sidharth's Compton phase objects \ct{50,51}.
Then the electroweak vacuum again presents the non-differentiable
manifold, and again we have to consider the non-commutative
geometry.

Kirzhnits \ct{Kir} and Linde \ct{1KL}  were first who considered
the analogy between the Higgs mechanism and superconductivity, and
argued that the SM (SU(2)-doublet) Higgs field condensate $v_1 =
\langle H\rangle \approx 246$ GeV disappears at high temperatures, leading to
the symmetry restoration. As a result, at high temperatures $T >
T_c$ all fermions and bosons are massless. These conclusions were
confirmed, and the critical temperature was estimated (see review
by A.~Linde \ct{Lin}).

 At the early stage, the Universe was very hot, but
then it began to cool down. Black-holes-monopoles (as bubbles of
the vapour in the boiling water) began to disappear. The
temperature dependent part of the energy density died away. In
that case, only the vacuum energy will survive. Since this is a
constant, the Universe expands exponentially, and an exponentially
expanding Universe leads to the inflation (see review \ct{1Lin}).
While the Universe was expanding exponentially, so it was cooling
exponentially. This scenario was called {\bf supercooling in the
false vacuum.} When the temperature reached the critical value
$T_c$, the Higgs mechanism of the SM created a new condensate
$\phi_{min 1}$, and the vacuum became similar to superconductor,
in which the topological defects are magnetic vortices. The energy
of black-holes is released as particles, which were created during
the radiation era of the Universe, and all these particles
(quarks, leptons, vector bosons) acquired their masses $m_i$
through the Yukawa coupling mechanism $Y_f \bar \psi_f\psi_f\phi$.
Therefore, they acquired the Compton wavelength,
$\lambda_i=\hbar/m_ic$. Then according to the Sidharth's theory of
the cosmological constant, in the EW-vacuum we again have
lattice-like structures formed by bosons and fermions, and the
lattice parameters ``$l_i$" are equal to the Compton wavelengths:
$l_i =  \lambda_i = \hbar/m_ic$.

As it was shown in Ref.\ct{38}, the Planck scale vacuum energy density
(with the VEV $v_2$) is equal to:
\be   \rho_{vac}({\rm at\,\, Planck\,\, scale}) =  \rho_{ZPF}({\rm
at\,\, Planck\,\, scale}) - \rho_{black\,\, holes}^{(NC)}\approx 0,
\lb{5pt} \ee
and the EW-vacuum gives:
\be   \rho_{vac}({\rm at\,\, EW\,\, scale}) =  \rho_{ZPF}({\rm at\,\,
EW\,\, scale}) - \rho_{vortex\,\, contr.}^{(NC)} -\rho_{boson\,\,
fields}^{(NC)}
   + \rho_{fermion\,\, fields}^{(NC)} \approx 0. \lb{6pt} \ee
In the above equations ``NC" means the ``non-commutativity" and
``ZPF" means ``zero point fields".

Assuming by example that hedgehogs form a hypercubic lattice with
lattice parameter $l=\lambda_{Pl}$, we have the negative energy
density of such a lattice equal to:
\be    \rho_{lat} \simeq - M_{BH}M_{Pl}^3.  \lb{7pt} \ee
If this energy density of the hedgehogs lattice compensates the
Einstein's vacuum energy (\ref{44}), we have the following
equation:
\be   \frac{\lambda}{4}v^4 \approx |M_{BH}|M_{Pl}^3,
                                     \lb{8pt} \ee
Using the estimation (\ref{38}), we obtain:
\be \frac 32 M_{Pl}^4 \approx |M_{BH}|M_{Pl}^3, \lb{9pt} \ee
or
\be |M_{BH}| = \frac 32 M_{Pl}\approx 3.65\times
10^{18}\,\, {\rm{GeV}}. \lb{10pt} \ee
Therefore hedgehogs have a huge mass of order of the Planck mass.
Eq.(\ref{20m}) predicts a radius $\delta$ of the hedgehog's core:
\be \delta \approx \frac{|M_{BH}|}{8\pi v^2}\approx
\left(\frac{128\pi}{3} M_{Pl}\right)^{-1}\sim 10^{-21}\,\, {\rm{GeV}}^{-1}.              \lb{11pt} \ee

\subsection{Stability of the EW vacuum}

Here we emphasize that due to the energy conservation law, the
vacuum density before the phase transition (for $T > T_c$) is
equal to the vacuum density after the phase transition (for $T <
T_c$), therefore we have:
\be  \rho_{vac}({\rm at\,\, Planck\,\, scale}) = \rho_{vac}({\rm at\,\,
EW\,\, scale}).     \lb{12pt} \ee
The analogous link between the Planck scale phase and EW phase was
considered in the paper \ct{50}. It was shown that the vacuum
energy density (DE) is described by the different contributions to
the Planck and EW scale phases. This difference is a result of the
phase transition. However, the vacuum energy densities (DE) of
both vacua are equal, and we have a link between gravitation and
electromagnetism via the Dark Energy. According to the last
equation (\ref{12pt}), we see that if $\rho_{vac}$ (at the Planck
scale) is almost zero, then $\rho_{vac}$ (at EW scale) also is
almost zero, and we have a triumph of the Multiple Point
Principle: we have two degenerate vacua with almost zero vacuum
energy density. Almost zero cosmological constants are equal: $$ \Lambda_1
= \Lambda_2\approx 0, $$ where $\Lambda_i$ is a cosmological
constant for $i$-vacuum with VEV $v_i$ (here $i=1,2$).

Now we see that we have obtained a very important result: our
vacuum, in which we live, is stable. The Planck scale vacuum
cannot be negative: $V_{eff}(min_1) = V_{eff}(min_2)$ exactly.

\section{Hedgehogs in the Wilson loops and the phase transition in the
$SU(2)$ Yang-Mills theory}

The authors of Ref.~\ct{BCK} investigated the gauge-invariant
hedgehog-like structures in the Wilson loops of the $SU(2)$
Yang-Mills theory. In this model the triplet Higgs field
$\hat{\Phi}=\frac 12 \Phi^a\sigma^a$ vanishes at the centre of the
monopole $x = x_0$: $\Phi(x_0) = 0$, and has a generic hedgehog
structure in the spatial vicinity of this monopole.

In the Yang-Mills theory, a hedgehog structure can be entirely
defined in terms of Wilson-loop variables \ct{C}. In general, we
consider an untraced Wilson loop, beginning and ending at the
point $x_0$ on the closed loop $C$:
\be  W_C(x_0) = P \exp ig \oint_C dx_{\mu}{\hat{A}_{\mu}}.
\lb{23w}  \ee
To improve the analogy with the triplet Higgs field  $\hat{\Phi}$, we subtract
the singlet part from $W_C(x_0)$:
\be \hat{\Gamma}_C(x_0) = W_C(x_0) - {\bf 1}\cdot \frac 12
TrW_C(x_0).  \lb{24w}  \ee
This is a traceless adjoint operator similar to the field
$\hat{\Phi}$. Associating the triplet part $\hat{\Gamma}_C(x_0)$
of Wilson loop $W_C(x_0)$ with the triplet Higgs field
$\hat{\Phi}$, we notice the following property: As the Higgs field
vanishes in all points $x$, belonging to the monopole trajectory,
similarly $\Gamma_C$ vanishes on the hedgehog loop $C$:
$$   W_C \in Z_2 \Leftrightarrow \Gamma_C = 0.$$

In conventional superconductivity \ct{48}, Abrikosov vortices are
singularities in the superconducting condensate (i.e., in the
Cooper-pair field). Abrikosov vortices are ``two-dimensional
hedgehogs" (see Ref.\ct{35}). In the core of the Abrikosov's
vortices, the superconductivity is broken, and the normal state is
restored. As temperature increases, the condensate weakens, and
nucleation of the vortices due to thermal fluctuations
strengthens. Thus, the higher the temperature is, the density of
the (thermal) vortices should be larger. It can be expected in the
YM theory that the density of hedgehog loops is also sensitive to
the phase transition.

The order parameter of the phase transition is the vacuum expectation
value (trace) of the Polyakov line:
\be  \hat{L}(x) = P \exp ig \int_0^{1/T} dx_4A_4(\vec x, x_4).
\lb{25w}  \ee
Here $T$ is a temperature and
VEV is $L = \frac 12 Tr\hat{L}$. Functional $\hat{L}(x)$, called
the {\bf thermal Wilson line}, is a basic variable in an effective
theory, which describes the properties of the finite-temperature
phase transition of the system. In the confinement phase, the
expectation value of the Polyakov line is zero: $\langle L\rangle
= e^{-TF_q} = 0$, indicating that the free energy of a single
quark becomes infinite when $F_q \to \infty$. In the deconfinement
phase, the Polyakov line has a non-zero expectation value:
$\langle L\rangle  \neq 0$, and the quarks are no longer confined.

Considering lattice model of the $SU(2)$ Yang-Mills theory,
Belavin, Chernodub and Kozlov showed numerically that the density
of hedgehogs structures in the thermal Wilson-Polyakov lines is
very sensitive to the finite-temperature phase transition. The
hedgehog line density behaves like an order parameter: the density
is almost independent of the temperature in the confinement phase
and changes substantially as the system enters the deconfinement
phase. These authors obtained a very important result:
$\beta_{crit}\approx 2.5$, which shows that the critical
     temperature $T_c$, corresponding to the hedgehogs confinement, is smaller than the Planck scale
     value.

Indeed,
\be \beta = 1/g^2 = 1/(4\pi \alpha)= \frac{1}{T\lambda_{Pl}}.
                                      \lb{26w} \ee
Then the critical temperature is:
\be  
    T_c = \frac {M_{Pl}}{\beta_{crit}}\approx 0.4 M_{Pl}\approx 10^{18}\,\, {\rm{GeV}}.                                                \lb{27w}
\ee

\section{Threshold energy of the $SU(2)$-triplet Higgs bosons}

Eq.(\ref{26w}) also gives the critical value of the coupling
constant $g^2_{crit}$ of the $SU(2)$ Yang-Mills theory:
\be   g^2_{crit}\approx 0.4,   \lb{28w}  \ee
or
\be  \alpha_{crit}^{-1} \approx 4\pi\times 2.5 \approx 31.4.
\lb{29w}  \ee
The renormalization group equation (RGE) for $\alpha^{-1}(\mu)$
(see for example \ct{52} and references there) is given by the
following expression:
\be \alpha^{-1}(\mu) = {\alpha(M_t)}^{-1} + b t,  \lb{30w}  \ee
where $t = \ln(\mu/M_t)$, and $M_t\simeq 173.34$ GeV is the top
quark mass.

Usually RGE is a function of $x$: $x=\log_{10}\mu$. Then
\be   t = \ln\left(\frac{10^x}{M_t}\right) =  x\ln{10} - \ln
M_t\approx 2.30x - 5.16. \lb{31w} \ee
For $SU(2)$-gauge theory $b\approx 19/12\pi$ and
$\alpha_2^{-1}(M_t)\approx 29.4\pm 0.02$, and we obtain the
following RGE equation \ct{52}:
\be \alpha_2^{-1}(x) \approx 29.4 + 0.504 (2.30x - 5.16).
                                        \lb{32w} \ee
Then we can calculate $x_{crit}$ using the following result:
\be \alpha_{crit}^{-1} \approx 31.4 = 29.4 + 1.16x_{crit} - 2.60,
                                        \lb{33w} \ee
which gives:
$$  x_{crit}\sim 4,$$  or $$ \mu_{crit} \sim 10^4 \,\,{\rm{GeV}}.$$ This result means that
the hedgehog's confinement happens at energy of 10 TeV, which is a
threshold energy of the production of a pair of the $SU(2)$-triplet
Higgs bosons $\Phi^a$:
 \be    E_{threshold}\sim 10^4 \,\, {\rm{GeV}} = 10 \,\, {\rm{TeV}}.
                         \lb{34w} \ee
At this energy we can expect to see at LHC the production of the
triplet Higgs particles with mass $\sim 5$ TeV. If we
assume that in the region $E > E_{threshold}$ the effective Higgs
potential has an interaction between the triplet field $\Phi^a$
and Higgs doublet $H^{\alpha}$ (here $a = 1,2,3$ and $\alpha =
1,2$), then we have such an effective Higgs potential with two
Higgs fields: $SU(2)$-triplet $\Phi^a$ and $SU(2)$-doublet $H$:
\bea V_{eff} &=& \lambda_{h,\, eff}(h)\left(|\Phi|^2 - v_2^2\right)^2 +
\lambda_{H,\,eff}(H)\left(|H|^2 - v_1^2\right)^2\nonumber\\&&+
\lambda_{hH,\,eff}(h,H)\left(|\Phi|^2 - v_2^2\right)\left(|H|^2 - v_1^2\right) +
\Lambda.
 \lb{35w} \eea
At $T=T_c$, we have the phase transition in the Universe when the
electroweak spontaneous breakdown of symmetry $SU(2)_L \times
U(1)_Y \to U(1)_{el.mag}$ creates new topological defects of the EW
vacuum: the Abrikosov-Nielsen-Olesen closed magnetic vortices
(``ANO strings") of an Abelian Higgs model \ct{48,49} and
point-like Compton phase objects \ct{50,51,SD}. Therefore below
energy $E=E_{threshold}$ we have the following effective Higgs
potential:
\be V^{(1)}_{eff} = \lambda_{H,\,eff}(H)\left(|H|^2 - v_1^2\right)^2 +
\Lambda,
 \lb{36w} \ee
which has the low-energy first vacuum with the VEV $v_1$.

Here it is necessary to comment that our Gravi-Weak unification
described in Section 4 is not valid exactly due to the presence of
a mixing term in the effective Higgs potential $V_{eff}$.
This unification is not correct if the mixing coupling constant
$\lambda_{hH,\,eff}$ is not very small and negligible. The
hedgehog's parameters obtained in Sections 4 and 5 are
approximately valid if $\lambda_{hH,\,eff} \ll 1$. In this paper we
assume that this coupling $\lambda_{hH,\,eff}$ is negligibly
small.

A cosmological constant $\Lambda$ in Eqs.~(\ref{35w}) and
(\ref{36w}) is given by the tiny value of DE (see Eq.~(\ref{3})).

\section{The Higgs mass and vacuum stability/metastability in
the Standard Model}

As it was mentioned in Section 2, assuming the existence of two
degenerate vacua in the SM (the first Electroweak vacuum  and the
second Planck scale one), Froggatt and Nielsen predicted the
top-quark and Higgs boson masses: $M_t = 173 \pm 5$ GeV and $M_H =
135 \pm 10$ GeV \ct{FN}. Their prediction for the mass of the SM
$SU(2)$-doublet Higgs boson was improved in Ref.~\ct{55} by
calculations of the two-loop radiative corrections to the
effective Higgs potential $V_{eff}(H)$ (here $H^2 \equiv
\phi^\dag\phi)$). The prediction of Ref.~\ct{55}: $M_H = 129 \pm 2$
GeV provided the possibility of the theoretical explanation of the
value $M_H \simeq 125.7$ GeV observed at LHC.

The authors of reference \ct{56} extrapolated the SM parameters up
to the high (Planck) energies with full 3-loop NNLO RGE precision.
From Degrassi et al. calculation \ct{55}, the effective Higgs
field potential $V_{eff}(H)$ has a minimum, which slightly goes
under zero, so that the present EW-vacuum is unstable for the
experimental Higgs mass $M_H \simeq 125.09 \pm 0.24$ GeV, while
the value that would have made the second minimum $v_2$ just
degenerate with the present vacuum $v_1$ would be rather $m_H
\simeq 129.4$ GeV.

A theory of a single scalar field is given by the effective
potential $V_{eff}(\phi_c)$ which is a function of the classical
field $\phi_c$. In the loop expansion $V_{eff}$ is given by a
series:
\be V_{eff} = V (0) + \Sigma_{n=1}V^{(n)},    \lb{29n} \ee
where $V (0)$ is the tree-level potential of the SM:
\be V (0) = - \frac 12 m_H^2\phi^2 + \frac 14\lambda_H \phi^4.
\lb{30n} \ee
The vast majority of the available experimental data is consistent
with the SM predictions. No sign of new physics has been detected.
Until now there is no evidence for the existence of any particles
other than those of the SM, or bound states composed of other
particles. All accelerator physics seems to fit well with the SM,
except for neutrino oscillations. These results caused a keen
interest in the possibility of the emergence of new physics only at very
high (Planck scale) energies and generated a great attention to
the problem of the vacuum stability: whether the EW-vacuum is
stable, unstable, or metastable. A largely explored scenarios
assume that new physics comes only at the Planck scale $M_{Pl} =
1.22 \times 10^{19}$ GeV. According to these scenarios, we need
the knowledge of the Higgs effective potential $V_{eff}(\phi)$ at
very high values of $\phi$.

The loop corrections give the $V_{eff}$ with values of
   $\phi$, which are much larger than $v_1\approx 246$
   GeV. The effective Higgs potential develops a
new minimum at $v_2 \gg  v_1$. The position of the second minimum
depends on the SM parameters, especially on the top and Higgs
masses, $M_t$ and $M_H$. This $V_{eff}(min2)$ can be higher or
lower than the $V_{eff}(min1)$ showing a stable EW vacuum (in the
first case), or metastable one (in the second case). The red solid
line of Fig.~2 by Degrassi et al. shows the running of the
$\lambda_{H,eff}(\phi)$ for $M_H \simeq 125.7$ GeV and $M_t \simeq
171.43$ GeV, which just corresponds to the stability line, that
is, to the stable EW-vacuum. In this case the minimum of the
$V_{eff}(H)$ exists at the $\phi = \phi_0 \sim 10^{18}$ GeV, where
according to MPP:
$$   \lambda_{H,eff}(\phi_0) = \beta(\lambda_{H,eff}(\phi_0)) = 0.$$
Unfortunately, according to Refs.~\ct{55,56}, this case does not
correspond to the current experimental values.

In Fig.~2 blue lines (thick and dashed) present the RG evolution
of $\lambda_H(\mu)$ for current experimental values $M_H \simeq
125.7$ GeV and $M_t\simeq 173.34$ GeV. The thick blue line
corresponds to the central value of $\alpha_s = 0.1184$ and dashed
blue lines correspond to its errors equal to $\pm 0.0007$. We see
that absolute stability of the Higgs potential is excluded by at
98\% C.L. for $M_H < 126$ GeV. Fig.~2 shows that asymptotically
$\lambda_H(\mu)$  does not reach zero but approaches to the
negative value:
\be   \lambda_H \to - 0.01\pm 0.002, \lb{31n}  \ee
indicating the metastability of the EW vacuum. According to the
paper \ct{55}, the stability line is given in Fig.~2 by the red
thick line and corresponds to $M_H = 129.4 \pm 1.8$ GeV. We see
that the current experimental values of $M_H$ and $M_t$ show the
metastability of the present EW-vacuum of the Universe, and this
result means that the MPP law is not exact.

\section{A new physics in the SM}

Can the MPP be exact due to the corrections from hedgehogs'
contributions? We think that it is possible.

If we assume that in the region $E > E_{threshold}$ the effective
Higgs potential contains not only the $SU(2)$-triplet field
$\Phi^a$, but also the $SU(2)$-doublet Higgs field $H^{\alpha}$
(where $a = 1,2,3$ and $\alpha = 1,2$), then there exists an
interaction (mixing term) between these two Higgs fields
as it was shown in Eq.~(\ref{35w}).
Of course, the effective Higgs self-interaction coupling constant
$\lambda_{H,\,\,eff}(\mu)$ is a running function presenting loop
corrections to the Higgs mass $M_H$, which arise from the Higgs
bosons $H$ $(\Delta\lambda_H(\mu))$ and from hedgehogs $h$
$(\delta\lambda_H(\mu))$:
\be \lambda_{H,eff}(\mu) = \frac{G_F}{\sqrt 2}M_H^2 +
\Delta\lambda_H(\mu) + \delta\lambda_H(\mu), \lb{33n}  \ee
where $G_F$ is the Fermi constant. The main contribution to the
correction $\delta\lambda_H(\mu)$, described by a series in the
mixing coupling constant $\lambda_{hH}$, is a term $\lambda_S$
given by the Feynman diagram of Fig.~3 containing the hedgehog $h$
in the loop:
\be \delta\lambda_H(\mu) = \Sigma_{n} c_n(\mu)\lambda_{hH}^{2n} =
\lambda_S(\mu) + .... \lb{34n}  \ee
Here the effective Higgs self-interaction coupling constant
$\lambda_{H,eff}(\mu)$ is equal to $\lambda_{eff}(\mu)$ considered
in Refs.~\ct{55,56}.

Our hedgehog is an extended object with a mass $M_h$ and radius
$R_h$, therefore it is easy to estimate $\lambda_S$ at high
energies $\mu > E_{threshold}$ by methods of Ref.~\ct{17}:
\be \lambda_S(\mu) \approx \frac
1{16\pi^2}\frac{\lambda_{hH}^2(\mu)}{(R_h M_h)^4}, \lb{35n}  \ee
where $\lambda_{hH}(\mu)$ is a running coupling constant of the
interaction of hedgehogs $h$ with the Higgs fields $H$ (see
Eq.~(\ref{35w})). In Eq.~(\ref{35}) parameters $M_h=|M_{BH}|$ and $R_h$ are the
running mass and radius of the hedgehog, respectively. According
to Eqs.~(\ref{20m}), (\ref{10pt}) and (\ref{11pt}), we have:
\be   M_h(\mu) = 8\pi v^2 \delta(\mu) \quad {\rm and} \quad
R_h(\mu) = \delta(\mu). \lb{36n} \ee
At high Planck scale energies, they are:
\be M_h \sim 10^{18}\,\, {\rm{GeV}},\quad
 R_h\sim 10^{-21}\,\, {\rm{GeV^{-1}}},     \lb{37n}  \ee
and
\be  R_h M_h \sim 10^{-3}.  \lb{38n}  \ee
As a result, asymptotically we have:
\be \lambda_S \sim \frac{{\lambda_{hH}}^2}{16\pi^2}10^{12}. \lb{39n}
\ee
If hedgehog parameter $\lambda_{hH}$ is:
\be \lambda_{hH}\sim 10^{-6}, \lb{40n} \ee
then
\be \lambda_S\sim 0.01, \lb{41n} \ee
and the hedgehogs' contribution transforms the metastable (blue)
curve of Fig.~2 into the stable (red) curve, and we have an exact
stability of the EW-vacuum and the accuracy of the MPP with two
degenerate vacua in the Universe.

A tiny value of the mixing coupling $\lambda_{hH}$, given by Eq.~(\ref{40n}),
    confirms a good accuracy of our calculations in the framework
    of the GWU model.
    Of course, the results obtained in our investigation depend on
details of the $f(R)$ gravity and Gravi-Weak unification model.
Nevertheless, we predict a production of triplet Higgs bosons at
LHC at energy scale $\sim 10$ TeV and the existence of two
degenerate, or almost degenerate vacua of our Universe provided by
the existence of black-holes-hedgehogs in the false Planck scale
vacuum.

\section{Conclusions}
\par\noindent
1. In this investigation, we were based on the discovery that a
cosmological constant of our Universe is extremely small, almost
zero, and assumed a new law of Nature which was named as a
Multiple Point Principle (MPP). The MPP postulates: {\it There are
two vacua in the SM with the same energy density, or cosmological
constant, and both cosmological constants are zero, or
approximately zero.} We considered the existence of the following
two degenerate vacua in the SM: a) the first Electroweak vacuum at
$v_1 = 246$ GeV, which is a ``true" vacuum, and b) the
second ``false" vacuum at the Planck scale with VEV $v_2 \sim
10^{18}$ GeV.
\par\noindent
2. The bubble, which we refer to as ``the false vacuum", is a de
Sitter space with its constant expansion rate $H_F$. The initial
radius of this bubble is close to the de Sitter horizon, which
corresponds to the Universe radius. The space-time inside the
bubble, which we refer to as ``the true vacuum", has the geometry
of an open FLRW universe.
\par\noindent
3. We investigated the topological structure of the universal vacua.
Different phase transitions, which were resulted during the
expansion of the early Universe after the Planck era, produced the
formation of the various kind of topological defects. The aim of
this investigation is the consideration of the hedgehog
configurations as defects in the false vacuum. We have obtained a
solution for a black-hole in the region which contains a global
monopole in the framework of the $f(R)$ gravity, where $f(R)$ is a
function of the Ricci scalar $R$. Here we have used the results of
the Gravi-Weak unification (GWU) model. The gravitational field,
isovector scalar $\Phi^a$ with $a = 1, 2, 3$, produced by a
spherically symmetric configuration in the scalar field theory, is
pointing radially: $\Phi^a$ is parallel to $\hat{r}$ -- the unit
vector in the radial direction. In this GWU approach, we obtained a
``hedgehog" solution (in Alexander Polyakov's terminology). We also
showed that this is a black-hole solution, corresponding to a
global monopole that has been ``swallowed" by a black-hole.
\par\noindent
4. We estimated all parameters of the Gravi-Weak unification
model, which gave the prediction of the Planck scale false vacuum
VEV equal to $v = 2\sqrt 2 M_{Pl}^{red}\approx 6.28\times
10^{18}$ GeV.
\par\noindent
5. We have shown, that the Planck scale Universe vacuum is
described by a non-differentiable space-time: by a foam of
black-holes, or by lattice-like structure, where sites are
black-holes with the ``hedgehog" monopoles inside them. This
manifold is described by a non-commutative geometry, leading to a
tiny value of cosmological constant $\Lambda\approx 0$.
\par\noindent
6. Taking into account that the phase transition from the ``false
vacuum" to the ``true vacuum" is a consequence of the electroweak
spontaneous breakdown of symmetry $SU(2)_L\times U(1)_Y \to
U(1)_{el.mag}$, we considered topological defects of EW-vacuum:
the Abrikosov-Nielsen-Olesen closed magnetic vortices (``ANO
strings") of the Abelian Higgs model and Sidharth's Compton phase
objects. We showed that the ``true vacuum" (EW-vacuum) again is
presented by the non-differentiable manifold with non-commutative
geometry leading to an almost zero cosmological constant.
\par\noindent
7. By solving the gravitational field equations we estimated the
black-hole-hedgehog's mass, radius and horizon radius are
$M_h\approx 3.65\times 10^{18}$ GeV,  $R_h \sim 10^{-21}$ GeV$^{-1}$
 and $r_h \approx
2.29R_h$ respectively.
\par\noindent
8. We considered that due to the energy conservation law, the
vacuum energy density before the phase transition is equal to the
vacuum energy density after the phase transition: $
\rho_{vac}({\rm at\,\, Planck\,\, scale}) = \rho_{vac}({\rm at\,\, EW\,\,
scale}).$ This result confirms the Multiple Point Principle: we
have two degenerate vacua $v_1$ and $v_2$ with an almost zero
vacuum energy density (cosmological constants). By these considerations we
confirmed the vacuum stability of the EW-vacuum, in which we live.
The Planck scale vacuum cannot be negative because of the exact
equality $V_{eff}(min_1) = V_{eff}(min_2)$.
\par\noindent
9. Hedgehogs in the Wilson loops of the $SU(2)$ Yang-Mills theory,
and phase transitions in this theory were investigated revising
the results of Refs.~\ct{BCK,C}. Using their lattice result for
the critical value of the temperature of hedgehog's confinement
phase: $\beta_{crit} \approx 2.5$, we predicted the production of
the $SU(2)$-triplet Higgs bosons at LHC at energy scale $\mu \sim
10$ TeV, providing a new physics in the SM.
\par\noindent
10. We considered an additional confirmation of the vacuum
stability and accuracy of the MPP taking into account that
hedgehog fields $\Phi^a$ produce a new physics at the scale $\sim
10$ TeV, and calculating at high energies the
contribution of the black-hole-hedgehog corrections to the
effective Higgs potential. This result essentially depends on the
hedgehog field parameters: mass, radius and mixing coupling
constant $\lambda_{hH}$ of the interaction of hedgehogs with the
SM doublet Higgs fields $H$.

\section*{Acknowledgments}
LVL greatly thanks to the B.M. Birla Science Centre (Hyderabad, India) and personally Prof.
B.G. Sidharth, for hospitality, collaboration and financial support. HBN wishes to
thank the Niels Bohr Institute for the status of professor emeritus and corresponding
support. CRD is thankful to Prof. D.I. Kazakov for support.

\newpage

\begin{figure}
\centering
\includegraphics[scale=0.55]{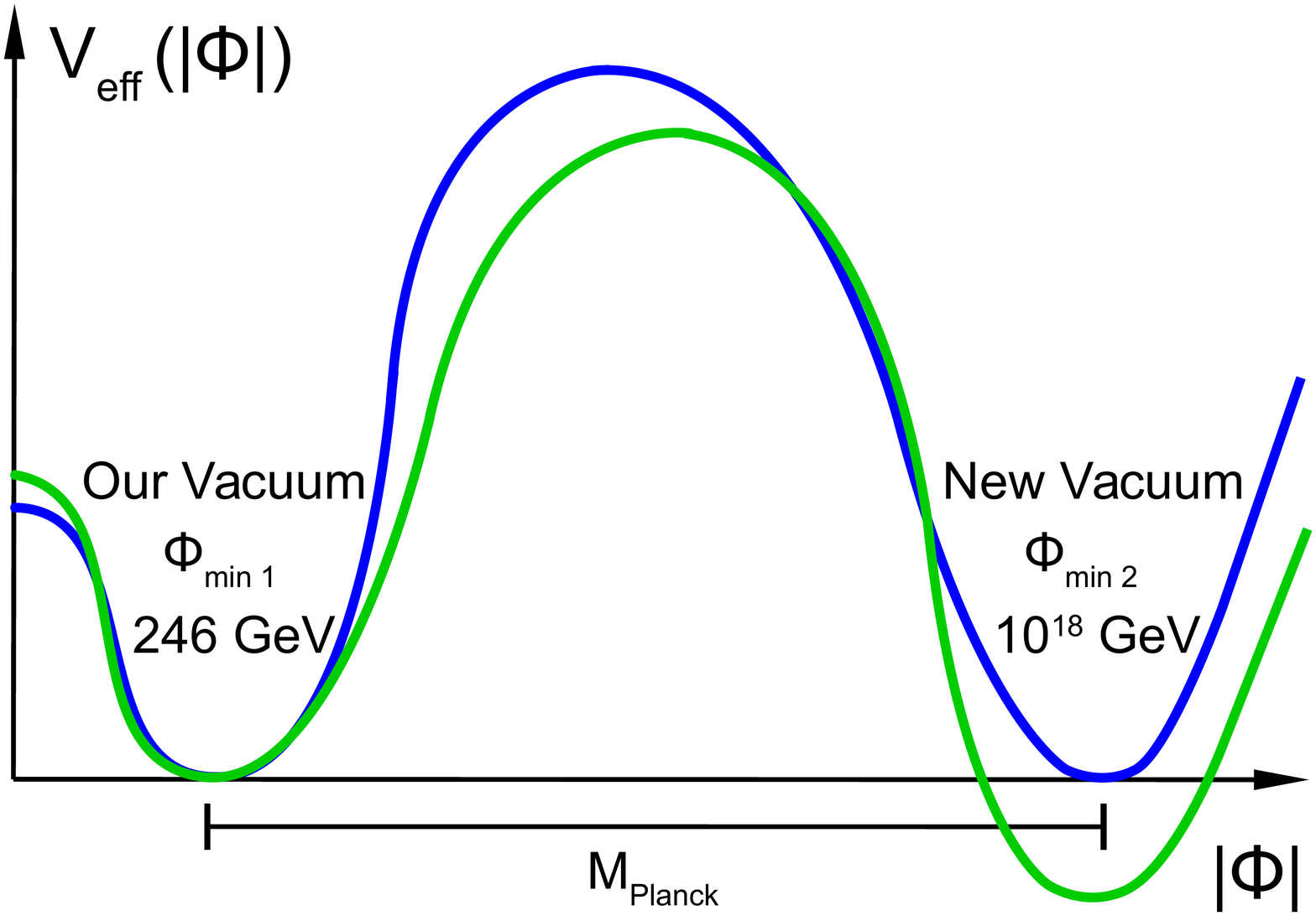}
\caption {Minima of the effective Higgs potential in the pure
Standard Model, which correspond to the first Electroweak ``true
vacuum'', and to the second Planck scale ``false vacuum''.}
\end{figure}

\newpage

\begin{figure}
\centering
\includegraphics[scale=0.65]{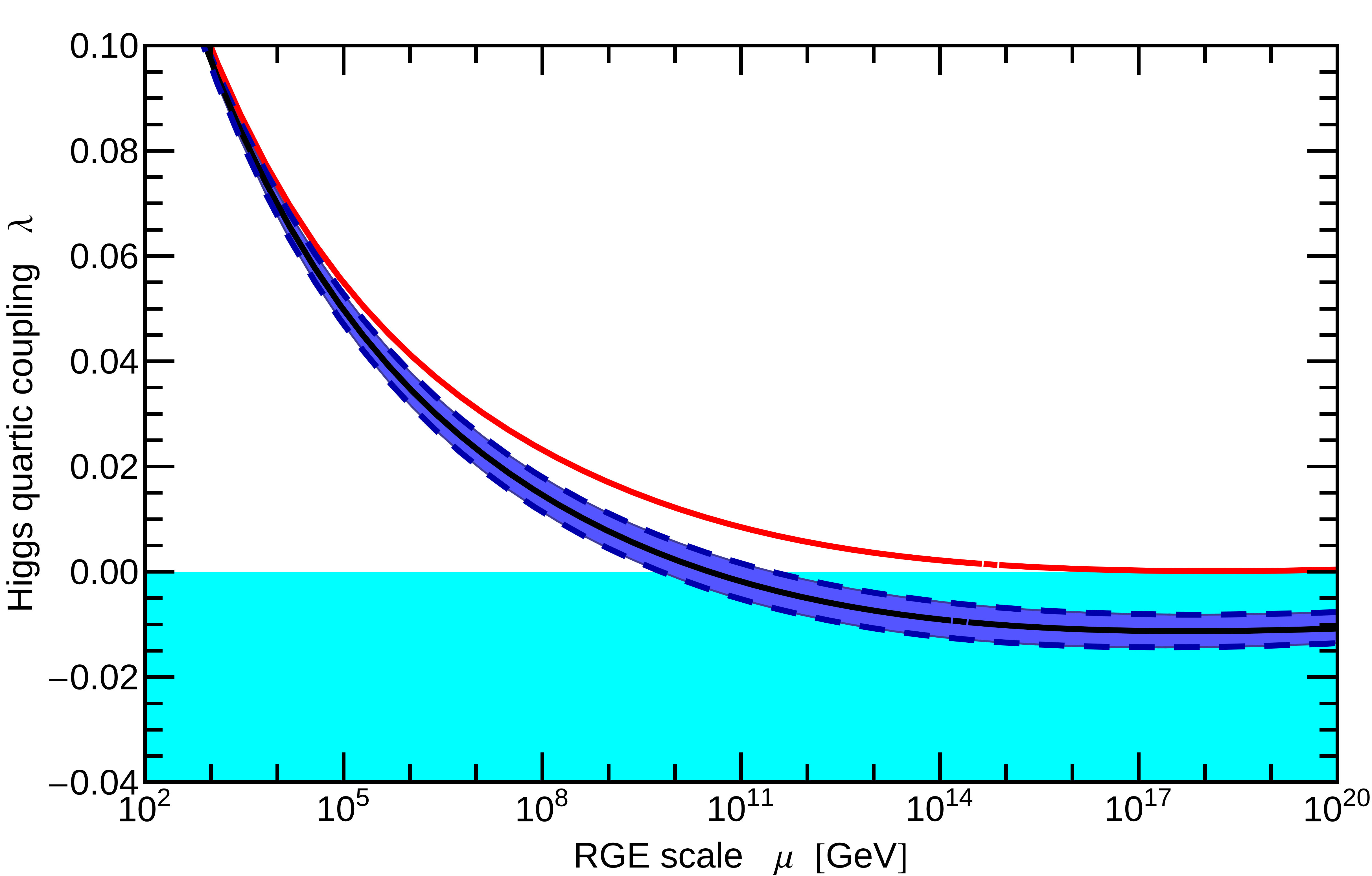}
\caption {The RG evolution of the Higgs selfcoupling $\lambda$ for
$M_t\simeq 173.34$ GeV and $\alpha_s = 0.1184$ given by $\pm
3\sigma$. Blue lines present metastability for current
experimental data, red (thick) line corresponds to the stability
of the EW vacuum.}
\end{figure}

\newpage

\begin{figure}
\centering
\includegraphics[scale=0.18]{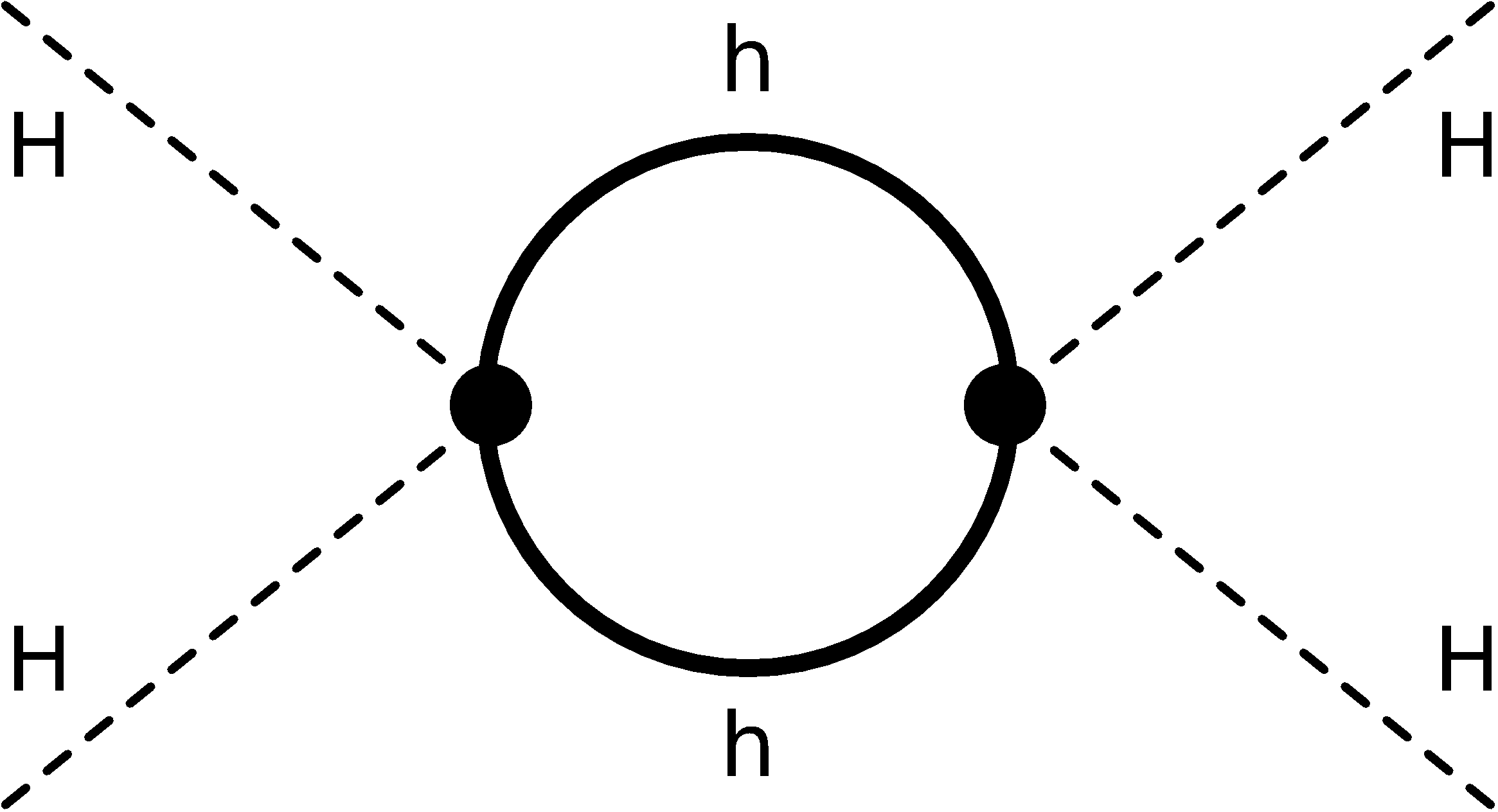}
\caption{The main Feynman diagram containing hedgehogs in the
loop, which corrects the effective Higgs mass.}
\end{figure}

\end{document}